# Understanding Organizational Approach towards End User Privacy

**Awanthika Rasanjalee Senarath**

Australian Centre for Cyber Security
School of Engineering and IT
University of New South Wales
Canberra, ACT
Email: a.senarath@student.unsw.edu.au

**Nalin Asanka Gamagedara Arachchilage**

Australian Centre for Cyber Security
School of Engineering and IT
University of New South Wales
Canberra, ACT
Email: nalin.asanka@adfa.edu.au

## Abstract

Abstract—End user privacy is a critical concern for all organizations that collect, process and store user data as a part of their business. Privacy concerned users, regulatory bodies and privacy experts continuously demand organizations provide users with privacy protection. Current research lacks an understanding of organizational characteristics that affect an organization's motivation towards user privacy. This has resulted in a "one solution fits all" approach, which is incapable of providing sustainable solutions for organizational issues related to user privacy. In this work, we have empirically investigated 40 diverse organizations on their motivations and approaches towards user privacy. Resources such as newspaper articles, privacy policies and internal privacy reports that display information about organizational motivations and approaches towards user privacy were used in the study. We could observe organizations to have two primary motivations to provide end users with privacy as voluntary driven inherent motivation, and risk driven compliance motivation. Building up on these findings we developed a taxonomy of organizational privacy approaches and further explored the taxonomy through limited exclusive interviews. With his work, we encourage authorities and scholars to understand organizational characteristics that define an organization's approach towards privacy, in order to effectively communicate regulations that enforce and encourage organizations to consider privacy within their business practices.

# 1  Introduction

Target, a popular retail chain, was accused of sending pregnancy catalogues to a 16 year old girl, whose pregnancy was not known to her parents (Forbes 2014). The data collection and processing methods Target adopted to enhance its marketing enabled it to collect and predict sensitive personal information of their customers such as, when they got divorced, when they got pregnant, and even when had a breakup (Forbes 2014). Such privacy invasive incidents being reported show how some organizations are increasingly investing in ways of collecting, storing and processing vast amounts of user data, without much concern on user privacy (Chan & Greenaway 2005), which have significant consequences on users whose data are compromised (Chan & Greenaway 2005).

With concerns for user privacy rising in the society, users (Sarvas & Frohlich 2011), the research community (Langheinrich 2001; Wright & De Hert 2012) and governments (Fromholz 2000) are demanding organizations prioritize privacy in their business practices. Such developments have made user privacy an increasingly important issue for organizations (Ginosar & Ariel 2017; Smith 1993; Julia 2009). Brunton & Nissenbaum (2017) claim, "In the digital economy, the real power is not held by individual consumers and citizens using their smart-phones and laptops to navigate the twists and turns of their lives, but by the large government and corporate entities who monitor them". However, due to the vast differences in the scale of operation, field of operation, nature of data stored and used, size, scope and revenues of different organizations, it is difficult to define the approach an organization should take to provide privacy to their users in a single model (Chan & Greenaway 2005). The "one solution fits all" approach taken so far in solving organizational privacy issues is not applicable anymore (Gürses & del Alamo 2016). Therefore, in order to understand, predict and solve organizational privacy concerns it is essential that the regulatory bodies and governments understand and acknowledge the organizations when they enforce laws and regulations. For this, here we attempt to empirically investigate the approaches taken by different organizations to address end user privacy requirements.

In this experiment, we studied 40 organizations that deal heavily with user data to develop, maintain and provide on-line applications to users, in-order to understand their approach towards privacy. Data was collected from organizational privacy policies, newspaper articles, previously published interviews and organizational reports on their data practices. Based on the results, we built a taxonomy of privacy protection approaches adopted by organizations towards end user privacy. We could observe four distinct approaches organizations take to provide privacy to their users characterized by their actions, expressions and communications of their priorities. Motivation is the key for people's actions, desires, and needs. Motivation is considered as the reason for behaviour, or what causes a person or an entity to want to repeat a behaviour (Elliot & Covington 2001). The four approaches we identified could be observed to be driven by two motivations described in psychology research. We could observe the approach an organization takes towards providing end user privacy was driven by either their inherent knowledge and business needs to attract customers, or as a risk management strategy to ensure compliance with regulations. We discuss these findings in detail together with the existing knowledge in psychology when we present our results. We then verified the taxonomy through a limited number of personal interviews with management personnel from a selected sub set of the organizations studied.

The paper is structured as below. The related work section extensively elaborate on research done so far in identifying organizational approaches towards privacy. The methodology section contains information on the study approach and this is followed by the results. We then discuss the results based on previous work in the field of privacy, and in the fields of business studies and organizational motivation, followed by our conclusions.

# 2  Related Work

Going through the limited research that exist on privacy as an organizational phenomenon, we could observe that most of them are skewed towards theory based interpretations rather than practical observations. These theoretical explanations attempt to frame organizational privacy behaviours into existing social science theories and then interpret the actions (Smith 1993). For example, Greenaway and Chan (2005) define a theoretical explanation on organizational privacy approaches through two theories. They claim that organizations either follow institutional theory due to external forces (legal and social), or resource-base-view theory considering user information as an organizational resource. They further go onto characterize organizations into two groups, based on their behaviour as "minimal privacy behaviours" where organizations demonstrate compliance to legal frameworks without much transparency on organizational strategies, and "enhanced behaviours" where organizations are more self-explanatory and informative about their privacy decisions. While these theories provide a strong background as a basis on which organizational privacy research could thrive on, they need to be interpreted with empirical evidence in order to establish their capacity to solve organizational privacy requirements, which is the focus of this study.

Anthonysamy et al. (2017) have identified four approaches towards privacy from an engineering perspective. Their classification tries to understand how privacy is implemented in a system considering it as an engineering requirement. On a similar perspective, Notario et al (2015) have identified two approaches to implement privacy protection during a software development process, as risk based and goal oriented. They define risk-based method as identifying threats to the system that might compromise the privacy of its end users and take measures to mitigate those risks in the development stage of the system. Goal oriented approach is defined as the approach where regulations and laws define principles the system must fulfil to provide data protection. Similarly, Van et al. (2003), in chapter 7 of their Handbook for Privacy and Privacy Enhancing Technologies, mention two categories, which could motivate organizations to perform privacy auditing, namely economic motive and social motive.

In contrast to the above theoretical approaches, Ginosar and Ariel (2017) in their study on the missing aspects of privacy research, identified web-site owners and management (essentially organizations) as an important stakeholder whose concerns, efforts and views has been missing from privacy research. In their analysis, similar to the theories put forward by Greenaway and Chan (2005), they claim that organizations are driven by institutional theory, where they create privacy policies as a response to external pressure, or as a resource base view, by identifying user information as an important resource to gain competitive advantage in their business. However, their study was a survey-based investigation. In a similar study, Schwaig et al. (2006) investigated the compliance to Fair Information Practices by the top 500 largest US corporations by total revenue. Their study was limited to compliance. In this work, we are investigating and understanding the efforts, concerns and attitudes organizations have as a whole towards, not only in providing privacy policies, but also in providing privacy protection through their applications to end users.

Our work includes an analysis of organizational policies, reports, and declared commitment towards privacy in order to understand their approach towards privacy. We did not have any prejudice as to how we believe organizations would approach user privacy or a motive to interpret organizational behaviours explicitly based on the theories mentioned above. The empirical evidence unveiled in this study describe, enhance and establish the theories described above, and helps the governments, researchers and the organizations themselves to better address organizational concerns on end user privacy requirements.

## 3    Study Methodology

Our goal in this study was to understand how different organizations approach end user privacy. For this, we conducted an empirical investigation of 40 diverse organizations. Below we describe how we selected the organizations and carried out the investigation in detail.

The first step of the study was to select organizations to study. For this, we first identified 5 categories of organizations that heavily deal with personal data of users as Electronic and Software Development, Banking and Insurance, Government, Telecommunication Service Providers and Online Sales and Service Providers. This list was compiled following an extensive study on recent breaching incidents through newspaper articles. In this preliminary study, we studied breaching incidents that appeared in newspapers articles available online in the last 5 years. Then, we selected 40 organizations overall representing all the above categories that differ significantly in organizational structure (open source, board controlled, privately owned), operational scale (international scale, locally based) and revenue (based on Forbes list of companies against their net worth) aiming to increase the validity and credibility of our taxonomy. We considered availability of data, ease of access of data and public interest when we selected organizations. The final selection consisted of 13 electronic and software development organizations, 8 banking and insurance, 2 government, 3 telecommunication service providers and 14 online sales and service providing organizations in this study.

The next step was to analyse the selected organizations in order to understand their approaches towards user privacy. Greenaway and Chan (2005) has previously defined differences in the communication of privacy among organizations that follow different approaches towards user privacy. Based on this work we used the content, wording, explanations and presentation of the privacy policy as a key element in understanding an organization's approach towards end user privacy. We used mixed data-collection method (Small 2011) for the analysis with privacy policies, newspaper articles, publicly available administrative reports, on-line resources and government reports and also interviews given by the organizations as our resources. We collected newspaper articles within the last 10 years concerning privacy incidents of the selected organizations, accessed on-line privacy policies of the organizations and downloaded publicly available materials in the organization's web site that relates to its approach and decisions towards end user privacy. We collected at least two newspaper, and not more than 8 articles on breaching incidents for each organization.

Triangulation method in qualitative research is an approach where different resources are incorporated in a study to enhance credibility and reliability of the results (Jick 1979). In the second step, going forward with the triangulation method, to challenge and further ground the results, we conducted exclusive structured interviews with technical and management personnel from the organizations. We selected and sent email invitations for interviews to 15 of the 40 organizations, of which 11 responded with expression of interest. All the organizations we selected for interviewing were either Australian based organizations, or those that had branches in Australia and hence, all the interviewees were based in Australia. Only 7 participants agreed to continue with the interview following the explanation of the interview questions. The random sample was chosen to represent each of the five categories of organizations we used in the first step. Interviewees were guaranteed that neither their personal profile, nor their company profile will be revealed in presentation of data gathered. We conducted the interviews over the phone. The interviewees were not compensated in any way for their participation, other than a verbal appreciation on their input as a professional in the field. Two of the participants had 2 to 5 years of experience in security and privacy and one participant had 5 to 10 years of experience. Four of the participants had more than 10 years of experience. The complete study design was approved by the ethic committee of the University of New South Wales.

### 3.1 Data Analysis:

Our study is based on the grounded theory approach (Corbin & Strauss 1990) and the outcome is based on the empirical evidence unveiled through the study (Corbin & Strauss 1990). Coding is a popular approach adopted in qualitative research as a reduction methodology for theory formation based on data gathered (Saldaña 2015). Similarly, in our approach we first summarized the privacy policy of each organization using open coding. Examples of the summarization codes we generated at this level are "we manage privacy risk", "we understand your privacy needs", "we participated in privacy sealing", "we strictly adhere to government laws", "we follow fair information practices / privacy by design". We used the other resources (internal reports, government reports) to interpret the abstract statements in the privacy policies, to assist the initial level of summarization. We then analysed this summary to identify the characteristics that defined that organization's approach towards user privacy. We categorized organizations that had similar codes and after several rounds of combining different categories, we ended up with four distinct organizational approaches towards end user privacy. They are government regulation compliance approach, government AND/OR self-regulations compliance approach, user focus approach and privacy education approach. We then performed axial coding on the common characteristics in each category to summarize the key factors.

In the grounded theory approach, considering literature is permitted in guiding data analysis (Suddaby 2006). Therefore, we made use of the theoretical contributions of defining organizational privacy approaches by Greenaway and Chan (2005) and Ginosar and Ariel (2017) to guide us in our coding process. We could see striking similarities in the codes we generated that clearly differentiated between the two approaches previously identified by Greenaway and Chan (2005). Their work on identifying and modelling organizational privacy as a resource based view and institutional theory unarguably became the backbone of our categorization. Thereby, building on this knowledge we could identify two motivational factors that drive the four organizational approaches towards end user privacy. We selectively re-coded some of the initial codes based on this knowledge. To demonstrate a more practical interpretation of our findings, we used the terms voluntary approach and risk based approach to identify the two motivations in our taxonomy. These terms appeared more applicable in defining the two groups due to some characteristics we found within the two categories which were not given significance in the original theory. For example, the element of risk as a catalyst in encouraging institutional theory based approach within organizations was disregarded in the theory by Greenaway. Further, the resource base view developing an inherent motivation was visible in our analysis, which we interpreted as voluntary motivation.

## 4  Study Results

In this section we present the results of the study we conducted to identify organizational approaches towards end use privacy. As we discussed in the data analysis section above, we identified four distinct organizational approaches towards end user privacy and two motivations that drive these approaches. This shows that organizations approach end user privacy in different ways, which confirms our previous claim that the "one solution fits all" approach taken in defining privacy laws and regulations does not adequately address organizational concerns on end-user privacy. The following table depicts the characteristics of the four organizational approaches towards end user privacy.

| Risk Induced, Gov. Regulation Compliance Based Approach (RISK-REG) | Risk Induced, Gov. Regulation AND/OR Self-Regulation Compliance Approach (RISK-SELF) | Voluntarily Induced Education Based Approach (VOL-EDU) | Voluntarily Induced User Focus Approach (VOL-USER) |
|---|---|---|---|
| Investigate gov. regulations due to potential risks for survival of their business | Declare their privacy policy with focus on potential privacy risks imposed on the organization. | Organizations that implement privacy because they think it is the right thing to do. | Provide end users with privacy they believe that fits best with user requirements. |
| Implement privacy according to govt. regulations through consultation of security and legal measures. | Whilst complying with government regulations conduct employee training. | Their privacy policies are mostly incomplete and complex and are not reflected in practice. | Frequently change and modify their privacy policies and release products to manipulate customer perception on privacy. |
| Mention that they are considering privacy as a part of government requirements. | In addition to gov. laws, adhere to best practices recommended by third parties due to risk. | Observed in large organizations that have their own R&D in privacy. | Observed in companies that build their business around user information. |
| Use terms such as "our internal decisions", "we are compliant" and "operational requirements" which may appear vague to a general user. | Use terms such as "our internal decisions"," your personal information" and "operational requirements" which may appear vague to a general user. | The solutions they come up with may or may not adhere completely to the rules declared by governments and authorities. | Conduct studies to figure out user requirements, or define user requirements through their experience. |
| Most privacy incidents happen due to unintentional mistakes and mismatched implementations of regulations. | Define organizational policies and regulations towards privacy which may be less than, equal to, greater than government laws. | Mention "in accordance with Privacy by Design" and "Following Fair Information Practices" in the privacy policy. | Use more specific terms as "we collect your location to show you our closest delivery outlet". |
| Operations are dependent on the region and country they operate in. | Obtain privacy certifications to display commitment. | Organizations that implement privacy because they think it is the right thing to do. | Privacy policies are well versed and readable and comprehensive. |
| Always disclose privacy breaching incidents according to regulations. | Mindful of how their competitors are adopting privacy and attempt to provide the same or better to their customers | Their privacy policies are mostly incomplete and complex and are not reflected in practice. | |

*Table 1 Organizational Approaches towards End User Privacy*

From our analysis, we identified 15 Organizations that followed RISK-REG approach, 10 following RISK-SELF approach, 7 with VOL-EDU approach and 8 following the VOL-USER approach. However, the knowledge of the existence of different organizational approaches towards end user privacy would be of no use unless we can predict and define how an organization would approach end user privacy depending on its characteristics. To differentiate solutions towards organizational concerns towards end user privacy based on their approaches, it is crucial that we find out what causes these difference and the reflections of the differences. To understand this, here we present our results analysing the privacy approaches against organizational characteristics in the following table.

## 4.1 Analysis Based on Organizational Characteristics

| Organization Type | RISK-REG | RISK-SELF | VOL-EDU | VOL-USER |
|---|---|---|---|---|
| Electronic and Software Manufacturing | - | - | 7 | 6 |
| Telecommunication Service Providers | 2 | 1 | - | - |
| Banking and Insurance | 6 | 2 | - | - |
| Online Sales and Service Providers | 7 | 5 | 1 | 1 |
| Government Service Providers | 2 | - | - | - |
| Business Model | RISK-REG | RISK-SELF | VOL-EDU | VOL-USER |
| Free Service Providers (ex : Google, Facebook Amazon) | 2 | 2 | 5 | 3 |
| Organizations selling services and products | 13 | 8 | 3 | 4 |
| Org. net worth (resource : Forbes 2016) | RISK-REG | RISK-SELF | VOL-EDU | VOL-USER |
| USD 20B or below | 8 | 3 | 2 | 2 |
| More than USD 20B up to 60B | 4 | 3 | 2 | 2 |
| More than USD 60B up to 100B | - | 2 | - | - |
| More than USD 100B | 3 | 2 | 3 | 4 |

*Table 2 Organizational Characteristics and their Approach towards End User Privacy*

## 4.2 Interview Results

The interview results further strengthened and enhanced the knowledge we determined by the desk investigation. One participant mentioned that he could "observe a significant improvement in the company's user base after integrating privacy concerns into the products, which motivated his organization to pay attention to privacy continuously". This organization was involved in social networking application development and the owner mentioned that user privacy concerns is a critical determinant when users adopt their applications. Another participant who is a manager in a financial organization suggested that the government should refine laws concerning the resale of client data for analytical purposes, as it would enable businesses to better perform and would also act as a deterrent to black market sales of user data. Another manager who represented an organization involved in security related software application development and management mentioned that most of the privacy decisions they take are based on their belief that it is the right thing to do, which demonstrates an education based approach. Another manager mentioned that following a huge security or privacy incident similar to the panama papers (The Guardian, Luke Harding 2016) gives them an incentive to be more concerned of privacy, showing organizational concerns towards privacy risk.

## 4.3 Limitations

Our study analysed 40 organizations, which might not be sufficient to provide a comprehensive statistical analysis of the results. The interviewees were all based in Australia. Studying more organizations with diverse business practices, would perhaps reveal more branches in the taxonomy.

# 5 Discussion

Our taxonomy depicted two branches of motivations that drive organizations to consider user privacy. Similar to Greenaway and Chan's (2005) theoretical model of organizational approaches towards privacy, the voluntary motivation we discovered was an "inside-out" approach where the organization was driven by internal concerns, considering user data as their resource, which required management.

The risk-based approach was an "outside-in" approach where the organization was driven by external factors to protect end user privacy. The taxonomy hence support the model by Greenaway and Chan (2005) and provide the background for it. For example, the study revealed that the organizations that were driven towards user privacy through risk had strong dependency on government regulations and regional infrastructure. On the other hand, strong dependency on user data within an organization's business model (social networking software providers) encourages a user focus approach towards privacy considering user data as their own resource. The following model explains the model we generated through the results.

| Dependency on Regional infrastructure | Low | Challenge State Regulations | Selective Adherence to State Regulations Based on Risk |
|---|---|---|---|
| | High | Adherence to State Regulations plus/minus Self-Regulations | Strict Adherence to State Regulations |
| | | High | Low |
| | | Dependency on User Data | |

*Figure 1 Modelling Privacy Approach*

Voluntary motivation was observed in 32.5% of the organizations studied. Even though this is low as a percentage, most of these organizations (Facebook, Apple) develop applications that are strongly connected with users' lifestyles. Hence, the impact these organizations have on user privacy is significant. Previous work has shown that some organizations approach privacy for business reasons, to attract a niche area of customers through declaration of privacy commitment, to promote the brand name and protect the market shares rather than due to risk (Asghari et al. 2016). These organizations publicly declare their commitment towards end user privacy, and use it as a marketing tool. Furthermore, previous work has suggested that organizations may consider privacy as a social responsibility and be attentive towards user privacy concerns (Straub Jr & Collins 1990). Social values, norms, and market demands in a society could act as an incentive encouraging an organization to voluntarily consider privacy in their business (Straub Jr & Collins 1990). This was the basis for the organizations (47%) that demonstrated motivation towards user privacy due to their knowledge that providing privacy protection to their users is the "right thing to do".

The other portion (53%) of voluntarily motivated organizations demonstrated a user focus approach towards privacy (VOL-USER). We found this approach to be similar to the interactive approach towards privacy, which was first coined by Gürses & del Alamo (2016) and defined as the methodology of capturing privacy matters that arise between peers or in a workplace due to the introduction of information systems, and improve user's agency with respect to privacy through socio-technical designs. In our study this was observed to be practiced only by social networking software providers (87%), possibly due to their wide interaction with billions of users and large scale operations with funds to conduct user interactive surveys.

## 5.1 Privacy, because users want it?

It has been shown that when users are concerned about privacy, they become reluctant to disclose information, which adversely affects the business of organizations that are dependent on user data (Ginosar & Ariel 2017; Smith 1993). Users who gain more knowledge on the business model of free service providers are realizing that if they are not paying for it, they are not the customer, but the product (Goodson 2012). Hence, merely complying with government regulations on data breach prevention and disclosure is not adequate for such organizations to convince their customers about the privacy protection they get (Fromholz 2000). Therefore, they take a more proactive, user focus approach towards privacy. However, organizations with user interactive privacy approach conduct research and experiments to not only understand user needs, but also to manipulate public perception on privacy. For example, Facebook introduced new privacy settings in 2010, stating that social norms towards privacy would evolve with time. Although they had to revoke and re-introduce a less complex version soon after, such experiments demonstrate the overall attitude social networking organizations in general have about user privacy expectations (The Guardian, 2013). These organizations undergo continuous legal penalties due to their mistakes with regards to providing users with privacy protection. For example, according to news, during the period of 2012-2014 alone, Facebook has paid more than US $30 million to settle law suits relating to privacy (ABC news 2014). In addition to that, they have been legally forced to abide by practice to consider privacy during development activities

(Electronic Privacy Information Center n.d.). However, legal penalties enforced by regulatory bodies have not been 100% effective in controlling voluntarily driven organizations due to ineffective communication and lack of understanding on the business practices of these organizations (Davies 2010). Due to the scale of operation and innovative business practices and requirements of these organizations, existing legal recommendations are perceived to be inadequate in serving their purpose.

## 5.2 Motivation and Privacy Risk

Extrinsic motivation is defined as doing an activity in order to attain some separable outcome, or to avoid a penalty (Carroll 1979) and would always be approached with the minimum possible effort to reach a pre-defined level of expectation (Olafsen et al. 2015). We could observe that the risk induced motivation identified in our taxonomy demonstrated characteristics similar to that of extrinsic motivation. The study revealed that organizations that demonstrated a risk based motivation towards privacy (67.5%) were mostly financial institutions, government organizations and telecommunication service providing organizations that were dependent on the government and the infrastructure in the region they operate in. Within the risk induced category 37.5% of the organizations demonstrated compliance towards government regulation (RISK-REG), which identifies the benchmark of privacy protection an organization should provide to an end user. Organizations in this category were observed to be motivated to provide users with privacy to enable them to operate their business in a country or a region. Additionally, we observed some organizations (40%) that complied with self-made regulations (RISK-SELF), which was either a sub set of govt. regulations or a more comprehensive exceeding set of regulations. Previous work has shown that risk based motivation is driven by stakeholder interests and competitor behaviours (Dusuki & Yusof 2016). Organizations that had risk induced privacy are observed to conduct risk identification processes to identify potential privacy impacts on stakeholders (Wright & De Hert 2012). We believe that this process results in a deep understanding of the system and its impacts on the end users. Organizations demonstrating an inclination to follow privacy regulations inevitably demand stronger involvement by the governing authorities to continue what they are doing in enforcing and defining regulations related to privacy (Voss 2017). However, to encourage organizational participation in adhering to these regulations, our study shows that it is necessary for the legal frameworks to interpret privacy as a risk, because, our taxonomy shows that organizations are motivated to adopt regulations due to risk.

With the monetary value of personal information in marketing and targeted advertising rising, the demand for mechanisms to control large organizations compromising user privacy against business motives is critical (Mai 2016). Nevertheless, voluntary motivation demonstrated by these organizations implies that they have an interest to understand and respond to user requirements in privacy. Voluntary motivation is similar to intrinsic motivation discussed in psychology. It is defined as doing of an activity for its inherent satisfactions rather than for some separable consequence (Carroll 1979). It is argued that intrinsic motivation is better in motivating a person towards a task compared to extrinsic motivation because the former is out of choice towards personal endorsement whereas the latter is a compliance due to an external control (Ryan & Deci 2000). Therefore, voluntary motivation, if properly monitored, could be used to shape the future of privacy research and development. We believe that governments and the research community, rather than attempting to bring organizations that have voluntary motivation towards privacy into legal frameworks, should focus on making use of the motivation they have together with their resources to redefine privacy to address their business motives. Such an approach would enable monitoring and directing the motivation they already have in a way that benefit both the users and the organization.

Interestingly, we could observe that some legal frameworks already encourage risk based compliance. For example, the latest General Data Protection Regulation of the European Union (GDPR), which is to be in action from 25th May 2018 (Voss 2017) enforces a risk based compliance. Legal compliance naturally enforces an element of risk on an organization that does not consider privacy during application development (Voss 2017). Not complying with existing laws could result in lawsuits that would damage an organization's reputation. Nevertheless, some regulations were observed to be focusing extensively on breaching, breach notification and compensation rather than proactive and preventive actions against risk (Garcia 2006; Romanosky et al. 2014; Fromholz 2000). Gürses & del Alamo (2016) points out that privacy is far more complex and vast than mere data breaching. Design flaws, lack of concern on privacy during business decision making, could bring consequences which have a higher impact on user privacy than data breaching incidents (Gürses & del Alamo 2016). Therefore, it is critical that the regulating bodies pay attention to how such unforeseen risks could be effectively conveyed to organizations. As our results revealed that a significant portion of organizations are motivated to embed privacy into their systems through risk (67.5%), we encourage the national and sectoral bodies that enforce privacy related regulations to improve their approach in the direction of highlighting privacy risks to induce motivation in organizations. Regulations such as the GDPR are showing signs of changes and initiating the required changes to integrate technological and engineering aspects into privacy. For such initiatives, the knowledge elicited here, which demonstrate how different organizations approach privacy as management strategy is important. Understanding how a particular sector of organization, for example banking and financial organizations adopting privacy due to risk suggest enforcing privacy rules through a risk based approach.

Anthonysamy et al. (2017) in their study on approaches to privacy implementation in systems, claim that consideration of privacy only being an early stage task, and the element of regulation and accountability to be positives of the compliance based approach. However, it is possible that the risks the organization perceive mismatch the actual risks perceived by the end users of a system. It has been shown that many developers feel that communicating with end users of a system is not necessary as they know what users want from a system (Caputo et al. 2016). Such attitudes could hinder an organization's capability to identify real risks as perceived by users. This may lead to mis-prioritization of risks and hence not deliver adequate privacy protection to the end users. Additionally, previous work has identified Compliance being limited to government legal documents, lack of concern for third party imposed risk on privacy, and non-adherence to the continuous changes in privacy requirements and functionalities are weaknesses of the Risk based approach (Anthonysamy et al. 2017). Similarly, our study further revealed the disjoint nature of privacy risk analysis and policy declaration from the technological and development practices to be a weakness. Interpretation of legal requirements and translating them into practice is a critical component that determines the compliance of organizations that approach privacy through regulations (Breaux & Antón 2008). Therefore, we suggest that legal frameworks make an effort to encourage an approach that demonstrate compliance.

### 5.3 Motivation at All Levels

A common characteristic that was evident in all motivational approaches we identified was that they are all top-down induced motivations. This essentially means that even though the organizational motivation towards privacy at the top level is either voluntary or compliance, for the ground level staff it is always compliance or obligation. It is understood that for better privacy implementation an organization's top management should enforce compliance by ground staff (Cavoukian et al. 2010). For example, Alge et al (2006) state that organizations should continuously monitor employees who may (un/willingly) manipulate the privacy practices exercised by a company. The recent incident at Uber, where employees were accused of spying on celebrity travel information in their systems (The Guardian 2016) is a good example for the need for such strict measures. However, we believe that if we can induce motivation for the ground level staff; development, quality assurance and legal teams to have voluntary motivation, the prevailing burdensome attitude towards privacy in organizations could be changed (Senarath et al. 2017). For example, technical organizations are moving towards flat hierarchal management strategies to encourage technological breakthroughs from ground level staff by giving them opportunity and authority (Brem & Wolfram 2017). We believe that applying the same for privacy practices could nudge ground level staff to cultivate a voluntary attitude towards privacy.

Our results suggest that the future direction in privacy research should be a mixed approach. Governments should make it a priority to regulate privacy policies and laws as more than half of the organizations studied were observed to be motivated to consider privacy in their business model due to risk induced compliance. Therefore, legal frameworks act as a strong incentive in not only encouraging the organizations to adopt privacy in their business practices, but also in setting the standards in privacy protection. However, our results also strengthen the claim by Van et al. (Van Blarkom et al. 2003) which states that self-regulation may not be sufficient when it comes to organizations, as they tend to adopt only what they find attractive. Therefore, inducing voluntary motivation at all levels within an organization hierarchy is essential to ensure privacy as an organizational practice. Further to that, when it comes to organizations that are heavily dependent on user data, which demonstrated a voluntary motivation towards privacy, the regulations need to be modified and customized. The results suggested that such organizations believed the existing laws to be outdated or non-considerate concerning their business practices and requirements. Hence, further attention is required as to how to allow mutual benefits for both users and organizations while using technology to preserve privacy.

## 6 Conclusion

In this work, we analysed organizational approaches to embed privacy into the systems they develop and use. We did a comprehensive study of 40 international scale organizations based on their declarations, public and government reports and interviews. The taxonomy we developed shows how organizational characteristics such as their business model, revenue, and the nature of business relate to the approach they take towards user privacy. These findings contribute to the knowledge that is required by regulatory bodies and governments to understand the organizations when they enforce privacy regulations. Further, our findings revealed that the regulatory bodies need to understand the technological advancements that drive organizations to change and challenge privacy laws. We believe that regulations enforced through such an understanding would positively influence adherence.